\documentclass{kluwer}    
\usepackage{graphicx}
\newdisplay{guess}{Conjecture}

\begin{document}                                                                                   
\def\teff{$T_{\rm eff}$~} 
\begin{article}
\begin{opening}         
\title{Binaries with total eclipses in the LMC: potential targets for
spectroscopy} 
\author{Pierre \surname{North}}  
\runningauthor{Pierre North}
\runningtitle{Totally eclipsing binaries in the LMC}
\institute{Laboratoire d'astrophysique, Ecole Polytechnique F\'ed\'erale
de Lausanne (EPFL), Observatoire, CH--1290 Sauverny, Switzerland}
\date{}

\begin{abstract}
35 Eclipsing binaries presenting unambiguous total eclipses were selected from
a subsample of the list of \inlinecite{W03}. The photometric elements are given
for the $I$ curve in DiA photometry, as well as approximate \teff and masses of
the components. The interest of these systems is stressed in view of
future spectroscopic observations.
\end{abstract}

\end{opening}           

\section{Introduction}  
As recalled by \inlinecite{WW01}, ``systems where the stars are completely
eclipsed are particularly important because they can provide robust measurements
of the ratio of radii'', $k$. Accurate determination of $k$ is indeed
a well known problem in partially eclipsed systems. An impressive demonstration
of this is provided by Fig.~3 of \inlinecite{G05}.
Here, I draw attention to a few tens of totally eclipsing systems in
the LMC, which would deserve spectroscopic observations for an orbit
determination, and possibly additional photometric ones for a more accurate
determination of radii and surface brightness ratio. Such systems will be
useful, not only for distance determination, but also for comparison with
stellar structure models.

\section{Sample, lightcurve solution and stellar parameters}
From the sample of Wyrzykowski et al. (2003) based on OGLE photometry, we have
selected a subsample of 510 binaries with $I_{\mathrm max}\leq 18.0$, a depth
of the secondary minimum $\geq 0.20$~mag and an EA type. This is the same sample
as that used by North \& Zahn (2004). All lightcurves were solved interactively
with the EBOP code, assuming a linear limb-darkening coefficient
$u_p = u_s = 0.18$ except in the few cases of clearly cool components.
Out of this sample, we selected visually 35 systems with clearly total
eclipses. The fundamental stellar parameters
were determined through interpolation in the evolutionary tracks of
\inlinecite{SMMS93} computed for a metal-content $Z=0.008$ typical of the LMC.
This was done as in \inlinecite{NZ04}, but without the
hypothesis of identical components and assuming
$(m-M)_0=18.5$. The relative radii, orbital period and surface brightness ratio
$J_s$ were used to constrain the solution. $J_s$ was
calibrated in terms of \teff ratio through the models of
\inlinecite{K79}. In addition, $E(B-V)$ and $A_V=3.1\;E(B-V)$ were determined
simultaneously, using the measured $B-V$ index, following \inlinecite{N04}.
 The condition that both components lie on the same isochrone was
not implemented, because of the presence of some post-mass exchange systems; for
the latter, the masses given are just those of single stars with same \teff and
luminosity, and therefore may be wrong. For main sequence systems lacking a $B-V$
index, stellar parameters were determined in a cruder way, assuming both
components lie on a $\log t=7.0$ isochrone. The \teff of cool giants in a few
systems were derived from the $B-V$ or the
$V-I$ index assuming $E(B-V)=0.143$,
while the masses were assumed identical to those of stars with same $M_V$ on the
isochrone.
Some parameters of the $I$ DIA lightcurve as well as \teff and mass of the primary
components are given in Table~1, a more complete version of which is available
at the site
\verb+http://obswww.unige.ch/~north/DEB/tot_param+.\\
The errors are the formal ones given by the EBOP code and give an idea
of the quality of the fit, though one has to keep in mind that some parameters
were kept fixed in the fit, so that the errors displayed are rather lower limits
to the real uncertainties.

\section{Discussion}
Many systems with total eclipses have also relatively shallow minima. Although
this might be due to 3rd light in some cases, this cannot explain all of them.
Many such systems are certainly pairs of main sequence stars with a relatively
small ratio of masses and radii. Among them, those for which $J_s\sim 1$ are
especially interesting: they are composed of two stars close to the turn-off,
with an evolved primary and an unevolved secondary.
The ratio of radii is often near $0.5$. Such systems allow to probe efficiently
the global metallicity and helium content of each component -- as far as
the stellar structure models can be trusted -- according to the method of
\inlinecite{RJTG00}. Since the metallicity of the LMC is less than half that of
the Sun, this can potentially improve the $\Delta Y/\Delta Z$ relation obtained
by Ribas et al. on the basis of Galactic systems. Adding totally eclipsing
systems of the SMC will further improve the determination of this relation,
even providing an independent estimate of the primordial He abundance through
extrapolation to zero metallicity.

However, the great interest of these systems has a price as regard to spectroscopic
observations: the small $k$ implies a small luminosity ratio --
especially in the blue ($\lambda$ domain of choice because of the
number of spectral lines) if the secondary is cooler than the primary --
so that high S/N spectra will be needed. The UVES instrument on the VLT, used
with a wide slit, may be appropriate. Another possibility would
be the use of FLAMES-GIRAFFE in the IFU ``low''
resolution mode ($R\sim 10000$), which would allow to observe a few systems
simultaneously.

\begin{figure} 
\centerline{\includegraphics[width=10cm]{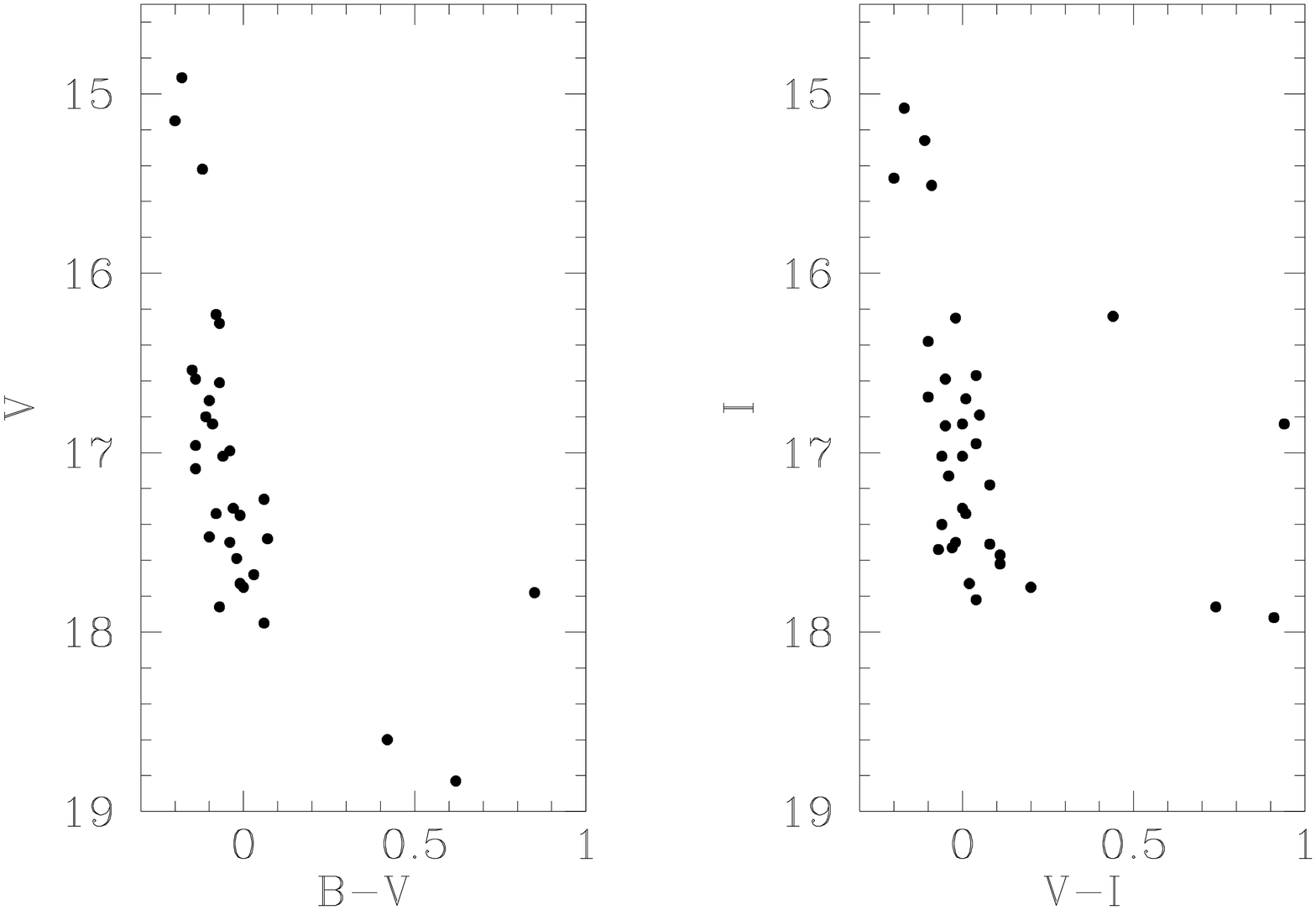}}
\caption{HR diagram of the totally eclipsing systems in the LMC.
Notice the three systems with giant components (a fourth one, in the
I vs V-I diagram, is an Algol-type system).}
\label{fig1}
\end{figure}

\begin{table*}
\caption[]{Totally eclipsing systems in the LMC. The errors on $r_p$, $k$ and
$e\cos\omega$ are given as the last digit(s) of the corresponding values.
The null uncertainties stand for parameters which were fixed
during the fit, though they were generally adjusted in a previous iteration. Eleven stars in the list are common to
\inlinecite{MP05}: W03 No 114, 362, 1148, 1291, 1520, 1551, 1748, 1880, 1996, 2279 and 2462. The
\teff of cool stars were estimated from $B-V$ with the calibration of \inlinecite{HK96} or from
$V-I$ with the calibration of \inlinecite{KH95} (their Table A5).}
\label{tab1}
\begin{tabular}{|crrrrrrrrrr|}
\hline
No&\multicolumn{1}{c}{$I_\mathrm{quad}$}
&\multicolumn{1}{c}{$P_\mathrm{orb}$}&
\multicolumn{1}{c}{$r_p$}&\multicolumn{1}{c}{$k$}&
\multicolumn{1}{c}{$e\cos\omega$}&
\multicolumn{1}{c}{$L_p$(I)}&\multicolumn{1}{c}{\teff$_p$}&
\multicolumn{1}{c}{$M_p$}&\multicolumn{1}{c}{\teff$_s$}&\multicolumn{1}{c|}{$M_s$}\\ \hline
 114&15.125&  2.98958& 0.258 2 & 0.537 04 &   0.0704 00&0.842& 34768 &  23.2 &  26930 &  11.1\\ 
 362&15.266&  3.37025& 0.306 1 & 0.733 03 &  -0.0093 03&0.682& 21999 &   9.6 &  20959 &   7.8\\ 
 398&16.862&  3.33782& 0.176 4 & 1.666 47 &   0.0031 00&0.563& 23133 &   7.6 &   8256 &   2.7\\ 
 406&18.037& 13.16448& 0.315 5 & 0.427 09 &   0.0000 00&0.965&  6387 &   6.1?&   4056 &   1.6? \\ 
 416&16.736&  5.16691& 0.175 3 & 0.473 05 &  -0.2392 09&0.829& 19066 &   6.6 &  17758 &   4.6\\ 
 460&16.618&  2.68062& 0.278 4 & 0.423 04 &   0.0005 00&0.866& 19924 &   7.2 &  18539 &   4.8\\ 
 670&16.604&  4.14459& 0.178 4 & 1.632 35 &   0.0058 00&0.561& 21405 &   7.0 &   8226 &   2.9\\ 
 810&16.743&  3.31516& 0.342 4 & 0.357 04 &   0.0061 00&0.896& 14032 &   4.9 &  13591 &   3.1\\ 
 841&17.861&  5.40797& 0.158 3 & 0.542 09 &  -0.0012 09&0.798& 12686 &   3.5 &  11453 &   2.5\\ 
1148&17.224&  2.16054& 0.262 2 & 0.653 05 &  -0.0341 09&0.741& 18293 &   5.7 &  16106 &   4.1\\ 
\hline
1208&17.602&  2.58542& 0.258 3 & 0.476 06 &   0.0192 18&0.867& 16308 &   5.0 &  12647 &   2.8\\ 
1212&18.034&  2.05110& 0.245 7 & 0.497 09 &   0.0535 23&0.858&  7348 &   5.9?&   6440 &   2.7? \\ 
1263&17.076&  6.12966& 0.175 2 & 0.556 05 &   0.1714 06&0.762& 14506 &   4.6 &  14514 &   3.6\\ 
1278&17.567&  7.42814& 0.168 3 & 0.413 07 &   0.0011 00&0.852& 11135 &   3.3 &  11313 &   2.4\\ 
1291&16.898&  5.23194& 0.184 3 & 0.504 05 &   0.1656 08&0.816& 17385 &   5.9 &  15840 &   4.0\\ 
1344&17.710&  4.14494& 0.233 5 & 0.505 08 &   0.0009 17&0.812& 11088 &   3.2 &  10558 &   2.3\\ 
1381&16.860& 79.17160& 0.126 2 & 0.526 07 &  -0.0016 10&0.819&  5552 &   8.6?&   5214 &   4.7? \\ 
1450&15.439&  2.72713& 0.330 1 & 0.605 02 &   0.0001 05&0.765& 31983?&  14.6?&  23557?&   8.2?\\ 
1469&17.334&  1.56400& 0.316 3 & 0.482 04 &   0.0006 00&0.889& 19480 &   6.1 &  13044 &   2.8\\ 
1520&16.082&  1.33832& 0.354 1 & 0.694 04 &   0.0000 00&0.720& 27348?&  10.8?&  21970?&   7.2?\\ 
\hline
1551&16.393&  1.53805& 0.349 2 & 0.778 05 &   0.0000 00&0.716& 22260 &   7.9 &  17536 &   5.2\\ 
1566&17.566&  3.23738& 0.223 6 & 0.450 09 &  -0.0256 13&0.864& 15696 &   4.7 &  13351 &   2.9\\ 
1675&17.574&  3.39105& 0.295 8 & 0.364 09 &   0.0088 00&0.886& 11436 &   3.5 &  11451 &   2.4\\ 
1748&15.544&  5.45728& 0.248 2 & 0.535 03 &   0.0254 06&0.800& 19860 &   8.4 &  18469 &   5.8\\ 
1880&17.202&  1.34524& 0.333 3 & 0.668 07 &  -0.0001 00&0.753& 18153 &   5.4 &  15305 &   3.8\\ 
1996&17.045&  1.82795& 0.258 3 & 0.535 05 &  -0.0037 12&0.850& 23050 &   8.1 &  16899 &   4.2\\ 
2009&17.570&  3.22935& 0.204 4 & 0.447 09 &  -0.1751 23&0.869& 18001 &   5.5 &  13873 &   3.0\\ 
2073&17.007&  2.89578& 0.303 5 & 0.386 06 &   0.0159 16&0.883& 14864 &   4.8 &  14108 &   3.1\\ 
2279&16.289&  3.31752& 0.257 2 & 0.691 07 &  -0.1347 09&0.680& 18293 &   6.5 &  18112 &   5.5\\ 
2289&17.789&  3.80450& 0.196 4 & 0.568 10 &   0.0001 15&0.769& 13681 &   3.9 &  13130 &   3.0\\ 
\hline
2380&17.374&  2.83324& 0.275 4 & 0.447 07 &   0.0939 20&0.852& 14296 &   4.4 &  13320 &   3.0\\ 
2462&16.767&  4.26120& 0.189 2 & 0.617 06 &   0.0993 10&0.766& 23378?&   8.1?&  17646?&   4.9?\\ 
2482&16.272&  8.07316& 0.177 4 & 1.547 39 &   0.0000 00&0.616&  8800?&   9.6?&   7050?&   8.0? \\ 
2533&17.748&  3.27882& 0.269 6 & 0.640 11 &  -0.0394 19&0.736& 10035 &   2.8 &   9311 &   2.2\\ 
2583&17.717&  2.07166& 0.266 4 & 0.567 09 &  -0.0346 20&0.788& 14909 &   4.2 &  13389 &   3.0\\ 
\hline
\end{tabular}
\end{table*}

\end{article}

\begin{thebibliography}{}

\bibitem[\protect\citeauthoryear{Gonz\'ales et al.}{2005}]{G05}
Gonz\'ales, J.~F., Ostrov, P., Morrell, N. and Minniti, D., 2005
\newblock {\em ApJ}, 624, 946
\bibitem[\protect\citeauthoryear{Hauck \& K\"unzli}{1996}]{HK96}
Hauck, B. and K\"unzli, M., 1996
\newblock {\em Baltic Astronomy}, 5, 303
\bibitem[\protect\citeauthoryear{Kenyon \& Hartmann}{1995}]{KH95}
Kenyon, S.~J. and Hartmann,L., 1995
\newblock {\em ApJS}, 101, 117
\bibitem[\protect\citeauthoryear{Kurucz}{1979}]{K79}
Kurucz, R.L., 1979
\newblock {\em ApJS}, 40, 1
\bibitem[\protect\citeauthoryear{Michalska \& Pigulski}{2005}]{MP05}
Michalska, G. and Pigulski, A., 2005
\newblock {\em A\&A}, 434, 89
\bibitem[\protect\citeauthoryear{North}{2004}]{N04}
North, P., 2004
\newblock {\em ASP Conf. Series}, 318, 273
\bibitem[\protect\citeauthoryear{North \& Zahn}{2004}]{NZ04}
North, P., Zahn, J.-P., 2004
\newblock {\em New Astron. Rev.}, 48, 741
\bibitem[\protect\citeauthoryear{Ribas et al.}{2000}]{RJTG00}
Ribas, I., Jordi, C., Torra, J., Gim\'enez, \'A., 2000
\newblock {\em MNRAS}, 313, 99
\bibitem[\protect\citeauthoryear{Schaerer et al.}{1993}]{SMMS93}
Schaerer, D., Meynet, G., Maeder, A., Schaller, G., 1993
\newblock {\em A\&AS}, 98, 523
\bibitem[\protect\citeauthoryear{Wyithe \& Wilson}{2001}]{WW01}
Wyithe, J.S.B, Wilson, R.E., 2001
\newblock {\em ApJ}, 559, 260
\bibitem[\protect\citeauthoryear{Wyrzykowski et al.}{2003}]{W03}
Wyrzykowski, L., Udalski, A., Kubiak, M. et al., 2003
\newblock {\em AcA}, 53, 1 (W03)

\end{thebibliography}
\end{document}